\documentstyle{article}

\def\ltap{\raisebox{-.4ex}{\rlap{$\sim$}} \raisebox{.4ex}{$<$}}

\newcommand{\Rsl}{{\not \! \!{R}}}
\newcommand{\rat}{\frac{m_t^2}{\tilde{m}^2}} 
\newcommand{\ratsq}{\frac{m_t^4}{\tilde{m}^4}} 

\begin{document}
\vspace*{-1in}
\newcommand{\Rslash}{{\not \! \!{R}}}
\renewcommand{\thefootnote}{\fnsymbol{footnote}}
\begin{flushright}
IFUP--TH 54/97\\
CUPP--97/3 \\
\texttt{hep-ph/9712245} 
\end{flushright}
\vskip 5pt
\begin{center}
{\Large{\bf New constraints on $R$-parity violation from $K$ and
$B$ systems}}
\vskip 25pt
{\bf Gautam Bhattacharyya $^{a,\!\!}$
\footnote{E-mail address: gautam@mail.cern.ch.~ Permanent address:
Saha Institute of Nuclear Physics, 1/AF Bidhan Nagar, Calcutta
700064, India.}}  and 
{\bf Amitava Raychaudhuri $^{b,\!\!}$
\footnote{E-mail address: raychaud@mail.cern.ch}} 
\vskip 10pt
$^a${\it Dipartimento di Fisica, Universit{\`a} di Pisa 
and INFN, Sezione di Pisa, I-56126 Pisa, Italy}\\
$^b${\it Department of Physics, University of Calcutta, 
92 Acharya Prafulla Chandra Road, \\ Calcutta 700009, India} 

\vskip 20pt
 
{\bf Abstract}
\end{center}

\begin{quotation}
{\small We have derived new upper limits on several products (two at a
time) of lepton number violating $\lambda'$-type couplings from the
consideration of $\Delta S = 2$ and $\Delta B = 2$ box graphs. Each
box contains one scalar lepton and one W-boson or one
charged-Higgs-boson as internal lines. Most of these bounds are more
stringent than previously obtained. Some of these product couplings
drive $K_L$ (and some other $B_d$) decays to two charged leptons at
enhanced rates.  Some of them can explain the rare $K^+ \rightarrow
\pi^+ \nu\bar{\nu}$ event recently observed at BNL.  \\ 
PACS number(s): 12.60.Jv, 13.20.Eb, 13.25.Hw, 14.80.Ly }
\end{quotation}

\vskip 20pt  

\setcounter{footnote}{0}
\renewcommand{\thefootnote}{\arabic{footnote}}

Unless one assumes that lepton-number ($L$) and baryon-number ($B$)
are conserved quantities, that are otherwise not ensured by gauge
invariance, supersymmetric theories with two Higgs doublets naturally
allow $R$-parity-violating ($\Rsl$) couplings \cite{rpar}. Defined as
$R = (-1)^{(3B+L+2S)}$ (where $S$ is the spin of the particle),
$R$-parity is a discrete symmetry under which all Standard Model (SM)
particles are even while their superpartners are odd. Even though any
concrete evidence of $L$- or/and $B$-violation is yet to be reported,
supersymmetry without $R$-parity has emerged as a fashionable area of
research of late \cite{Review}. When $R$-parity is violated, the
lightest supersymmetric particle does not remain stable and therefore
the canonical missing energy signature of supersymmetry search is no
longer valid. However, depending on the nature of $\Rslash$ couplings,
novel supersymmetry signatures ({\it e.g.}~multilepton or like-sign
dilepton final states \cite{DP}) emerge that could lead to the
discovery of supersymmetry in the present or future colliders. In
parallel with this, it is important to take a
stock of the extent to which those couplings are already constrained
from existing phenomenology. In this paper, we derive multitude of new
upper bounds on several combinations of the products of
$\lambda'$-type couplings (defined below), taken two at a time with
different flavour indices, that contribute to the $K_L$--$K_S$ mass
difference ($\Delta m_K$) or to the $B_q$--$\bar{B}_q$ ($q = d, s$)
mass differences ($\Delta m_{B_q}$) at one-loop level. We compare our
limits with the previous ones.  We also find enhanced rates for
$K_L\rightarrow \mu^+\mu^-$, $B_d \rightarrow \tau^+\tau^-$ and $K^+
\rightarrow \pi^+ \nu\bar{\nu}$, that proceed at tree level, in the
presence of some of those product couplings.

The most general Yukawa superpotential of an explicitly broken
$\Rslash$ supersymmetric theory is given by
\begin{equation}
 {\cal W}_{\not R}  =  {1\over 2}\lambda_{ijk} L_i L_j E^c_k
                     +  \lambda'_{ijk} L_i Q_j D^c_k 
                     +  {1\over 2}\lambda''_{ijk} U^c_i D^c_j D^c_k,  
\label{eq1}
\end{equation}
where $L_i$ and $Q_i $ are SU(2)-doublet lepton and quark superfields
respectively; $E^c_i, U^c_i, D^c_i$ are SU(2)-singlet charged lepton,
up- and down-quark superfields respectively; $\lambda_{ijk}$- and
$\lambda'_{ijk}$-types are $L$-violating while $\lambda''_{ijk}$-types
are $B$-violating Yukawa couplings.  $\lambda_{ijk}$ is antisymmetric
under the interchange of the first two family indices, while
$\lambda''_{ijk}$ is antisymmetric under the interchange of the last
two. Thus there could be 27 $\lambda'$-type and 9 each of $\lambda$-
and $\lambda''$-type couplings. Stringent constraints on the
individual couplings have been placed from the consideration of
$n$--$\bar{n}$ oscillation \cite{nnbar}, $\nu_e$-Majorana mass
\cite{numass}, neutrinoless double beta decay \cite{dbeta},
charged-current universality \cite{BGH}, $e$--$\mu$--$\tau$
universality \cite{BGH}, $\nu_\mu$ deep-inelastic scattering
\cite{BGH}, atomic parity violation \cite{BGH,Dreiner}, $\tau$-decays
\cite{BC,Bhatta}, $D$-decays \cite{BC}, $Z$-decays \cite{zdk,Bhatta},
$K^+$-decay \cite{AG} and $M^0$--$\bar{M}^0$ ($M \equiv K, B, D$)
mixing \cite{AG,Dreiner}. Products couplings (two at a time) have been
constrained by considering proton stability \cite{pdk}, neutrinoless
double beta decay \cite{BM}, meson mass differences and decays
\cite{BSB,CR,CW}, $\mu$--$e$ conversion \cite{BSB}, $\mu \rightarrow
e\gamma$ \cite{CH}, $b \rightarrow s \gamma$ \cite{CW}, $B$ decays
into two charged leptons \cite{Nardi,JKL} and $CP$-violation
\cite{cp}.

We first note that the SM contribution \cite{GL} to $\Delta m_K$ is
$\sim 2 \times 10^{-15}$ GeV and, on account of the large implicit
theoretical errors, it is believed to agree with the experimental
value ($\Delta m_K^{\rm exp} \simeq 3.5\times 10^{-15}$ GeV
\cite{pdg}) fairly well.  While placing bounds on any new physics that
contributes to $\Delta m_K$, we require that none of the new
contributions individually exceeds the experimental value. With this
view, we ignore the contribution coming purely from the Minimal
Supersymmetric Standard Model \cite{HKT} and restrict ourselves to
discussions of only $\Rslash$ contributions.  It has been argued in
\cite{AG} that one non-zero $\Rslash$ coupling in the weak basis of
fermions could lead to more such non-zero couplings in their mass
basis contributing to $\Delta m_K$. The bounds on the weak basis
couplings were estimated to be order 0.1 for 100 GeV scalar
exchanges. However, such bounds crucially depend on what assumption is
made during the basis rotation. In the present paper, we avoid
undergoing such a basis rotation and instead right-away assume {\it
two} non-zero $\Rslash$ couplings in their {\it mass basis} (in the
same notation) in each case.  Now we observe that the combinations
$\lambda'_{i21}.\lambda'_{i12}$ and $\lambda'_{i31}.\lambda'_{i13}$
contribute to $\Delta m_K$ and $\Delta m_{B_d}$ ($\Delta m_{B_d}^{\rm
exp} \simeq 3.3\times 10^{-13}$ GeV \cite{pdg}), respectively, at tree
level. These yield $\lambda'_{i21}.\lambda'_{i12} \ltap 1\times
10^{-9}$ and $\lambda'_{i31}.\lambda'_{i13} \ltap 8\times 10^{-8}$,
when the exchanged scalar has a mass of 100 GeV. However, there are
many other such product couplings that contribute to the $\Delta S =
2$ or $\Delta B = 2$ process {\it via} one-loop box graphs. In such
cases, all four vertices of a given box diagram (corresponding to
either $\Delta S = 2$ or $\Delta B = 2$) could be of the
$\lambda'$-type \cite{CW}. As an illustration, if we take two
$\lambda'_{i31}$- and two $\lambda'_{i32}$-vertices in a $\Delta S =
2$ box graph, the effective Hamiltonian could be expressed as
\begin{equation}
{\cal{H}}^{\Delta S = 2}_{4\lambda'} =
\frac{\left(\lambda'_{i31} \lambda'_{i32}\right)^2}{128\pi^2\tilde{m}^2}
{\cal{J}}(\rat) [\bar{d}\gamma_\mu (1+\gamma_5) s]^2, 
\label{eq2} 
\end{equation} 
where ${\cal{J}}(x) = (1+x)/(1-x)^2 + 2x\ln x/(1-x)^3$.  In the vacuum
saturation approximation, using $\langle K_0|[\bar{d}\gamma_\mu
(1+\gamma_5) s]^2|\bar{K}_0\rangle \simeq 4f_K^2 m_K/3$ and $\Delta
m_K = 2~{\rm Re}~(\langle K_0|{\cal{H}}|\bar{K}_0\rangle)$, we obtain
$\lambda'_{i31}.\lambda'_{i32} \ltap 2.8\times 10^{-3}$ for $\tilde{m}
\equiv m_{\tilde{L}_i} = 100$ GeV (also in all our subsequent
calculations, we take the mass of the exchanged scalar as 100
GeV). Throughout we use $m_t = 175$ GeV and $f_K = 150$ MeV. The
corresponding bounds when the internal fermions are instead the
$c$-quarks (or $u$-quarks) are $\lambda'_{i21}.\lambda'_{i22} ({\rm
or}~\lambda'_{i11}.\lambda'_{i12}) \ltap 1.2\times 10^{-3}$. On the
other hand, a similar computation of the $B_d$--$\bar{B}_d$ mixing
yields (with $f_{B_d} = 200$ MeV) $\lambda'_{i31}.\lambda'_{i33} \ltap
6.4\times 10^{-3}$ and $\lambda'_{i21}.\lambda'_{i23} ({\rm
or}~\lambda'_{i11}.\lambda'_{i13}) \ltap 2.7\times 10^{-3}$.

However, the main thrust of our analysis lies in computing another set
of $\Delta S = 2$ and $\Delta B = 2$ box graphs, hitherto overlooked,
in each of which there is one $\tilde{L}_i$ propagator between two
$\lambda'$-type vertices and one of $W^\pm$ (transverse $W$-boson),
$G^\pm$ (longitudinal $W$-boson in the 't Hooft - Feynman gauge) and
$H^\pm$ (charged Higgs) as the other non-fermionic
propagator\footnote{$\lambda''$-vertices with an internal $W$-boson
have been considered in \cite{CW}.}. The nature of the internal
fermion lines are decided by the flavour indices associated with the
two $\lambda'$-vertices. These diagrams are present in any $\Rslash$
supersymmetric theory, and as we will see, they serve to constrain
other flavour combinations of product couplings in addition to those
considered above. With $H^{\pm}$ (or $G^\pm$) as one scalar propagator
and, as always, $\tilde{L}_i$ as another between the
$\lambda'$-vertices, box graphs with $t$-quarks in internal lines
contribute more than those with $c$- or $u$-quarks, despite the
relatively larger Cabibbo-Kobayashi-Maskawa (CKM) suppressions
associated with the former. For light quarks ($c$ or $u$) as internal
fermions, box graphs with internal $W^\pm$ dominate over those with
internal $H^\pm$ or $G^\pm$, for the same choice of
$\lambda'$-couplings. Below we consider these possibilities case by
case:

\noindent {\it (i) Bounds from $\Delta m_K$}:~ First we choose the
same two $\lambda'$ as displayed in eq.~(\ref{eq2}), namely,
$\lambda'_{i31}$ and $\lambda'_{i32}$, as the two $\Rslash$ vertices
of a $\Delta S = 2$ box. Between the other two vertices in the 
box could flow either of $W^{\pm}$, $G^{\pm}$ and $H^{\pm}$. Evidently
$t$-quark propagates in internal fermion lines.  Assuming $m_{H^\pm} =
\tilde{m}$ (which is a reasonable assumption for an order of magnitude
estimate of the upper bounds of the $\Rslash$ couplings), the dominant
part of the effective Hamiltonian for this process could be written as
\begin{equation}
{\cal{H}}^{\Delta S = 2}_{2\lambda';t-t} \simeq \frac{G_F \lambda'_{i31}
\lambda'_{i32}}{16\pi^2\sqrt{2}} V_{ts}V_{td}^* \left[(\cot^2\beta + 1)
\ratsq {\cal{I}}(\rat) + {\cal{J}}(\rat)\right]
[\bar{d}(1-\gamma_5)s][\bar{d}(1+\gamma_5)s],    
\label{eq3}
\end{equation} 
where ${\cal{I}}(x) = -2/(1-x)^2 + (x+1)\ln x/(x-1)^3$ and $\cot\beta
= v_d/v_u$, the ratio of the vacuum expectation values of the two
Higgs bosons that are responsible for the generation of down- and
up-quark masses respectively. In eq.(\ref{eq3}), $H^{\pm}$- and
$G^{\pm}$-induced contributions carry an enhancement factor
($m_t^4/\tilde{m}^4$), the former involving $\cot^2\beta$; the last
term arises from $W^{\pm}$ propagation.  Within the vacuum saturation
approximation, employing $\langle
K_0|[\bar{d}(1+\gamma_5)s][\bar{d}(1-\gamma_5)s]|\bar{K}_0\rangle
\simeq 10f_K^2 m_K$, we obtain, for $\cot\beta = 1$,
$\lambda'_{i31}.\lambda'_{i32} \ltap 7.7\times 10^{-4}$. Noteworthy is
the point that even when $\cot\beta \ll 1$, the Goldstone contribution
holds the bounds at the same order of magnitude\footnote{In the
effective Hamiltonian, we have neglected a term proportional to
$\tan^2\beta$, arising from the charged Higgs coupling, that
multiplies light quark masses on external legs. Even for very large
$\tan\beta \sim 40$, the numerical contribution of such a term is
insignificant compared to the Goldstone contribution. This is true
even in the case of $B_s$--$\bar{B}_s$ mixing that we will discuss
subsequently.}.

An analogous process induced by $\lambda'_{i31}$ and $\lambda'_{i22}$
(and hence with $t$- and $c$-quarks as the two internal fermions)
leads to the effective Hamiltonian
\begin{equation}
{\cal{H}}^{\Delta S = 2}_{2\lambda';t-c} \simeq \frac{G_F \lambda'_{i31}
\lambda'_{i22}}{16\pi^2\sqrt{2}} V_{ts}V_{cd}^* \left[(\cot^2\beta + 1)
\frac{m_t^2 m_c^2}{\tilde{m}^2}{\cal{K}}(\rat) + {\cal{L}}(\rat)\right]
[\bar{d}(1-\gamma_5)s][\bar{d}(1+\gamma_5)s],    
\label{eq4}
\end{equation} 
where ${\cal{K}}(x) = 1/(x-1) - \ln x/(x-1)^2$ and ${\cal{L}}(x) = 1 -
x{\cal{K}}(x)$.  The following bounds are
$\lambda'_{i31}.\lambda'_{i22} \ltap 1.0\times 10^{-4}$. Similarly,
with $c$-quarks (or $u$-quarks) as internal fermions, we obtain the
bounds $\lambda'_{i21}.\lambda'_{i22} \ltap 1.4\times 10^{-6}$ and
$\lambda'_{i11}.\lambda'_{i12} \ltap 1.4\times 10^{-6}$. In the last
two cases, there are almost no CKM-suppressions and hence such tight
constraints. Bounds on other combinations are also obtained.
 
\noindent {\it (ii) Bounds from $\Delta m_B$}:~ The $\Delta B = 2$
processes are very similar to the $\Delta S = 2$ ones considered
above. Let us first consider $B_d$--$\bar{B}_d$ mixing with
$\lambda'_{i31}$ and $\lambda'_{i33}$ at the two $\Rslash$ vertices
({\it i.e.}~with $t$-quark in both internal fermion lines), and
$W^\pm$ or $G^\pm$ or $H^\pm$ propagating between the other two, as
before, in a given box graph. The effective Hamiltonian is exactly
analogous to the one in eq.~(\ref{eq3}), and we do not display it
here. However, in the present case, the CKM entries are $V_{tb}$ and
$V_{td}^*$ and the hadronic matrix element turns out to be $\langle
B_d|[\bar{d}(1+\gamma_5)b][\bar{d}(1-\gamma_5)b]|\bar{B}_d\rangle
\simeq 7f_{B_d}^2 m_{B_d}/6$. Again for $\cot\beta = 1$, we obtain
$\lambda'_{i31}.\lambda'_{i33} \ltap 1.3\times 10^{-3}$. We have also
derived upper bounds on other product couplings, corresponding to
heavy-light ($t-c, t-u$) or light-light ($c-c, c-u, u-u$) combinations
in the two internal fermion lines, that contribute to
$B_d$--$\bar{B}_d$ mixing\footnote{For our order of magnitude
estimate, we have neglected the imaginary parts that could arise when
both internal fermions are sufficiently light. Also all
$\lambda'$-couplings have been assumed to be real in the present
analysis.}. Analogous bounds from $B_s$--$\bar{B}_s$ mixing have also
been derived\footnote{We have used $f_{B_s} = 230$ MeV and $\Delta
m_{B_s}^{\rm exp} \simeq 3.9 \times 10^{-12}$ GeV (which is actually a
lower limit) \cite{pdg}.}.

We have listed all of our bounds in Table 1. While comparing our
bounds with the previous ones (the latter obtained by multiplying the
bounds on the individual couplings), it has to be borne in mind that,
keeping $\tilde{m}$ fixed, pushing $m_{H^\pm}$ to higher values indeed
weaken our limits but not the previous ones. However, the fact remains
that all existing bounds, that we compare with, have been derived
assuming a mass of 100 GeV for whichever scalar is exchanged, and that
way our assumption, that one more scalar (the charged Higgs boson) has
the same mass, is not unreasonable. It should be noted that we have
not indulged ourselves into computing the effects of QCD corrections
for our order of magnitude estimates.

Next we proceed to study the decays of neutral $K$- and $B$-mesons (in
particular, those of $K_L$ and $B_d$) into two charged leptons. In
$\Rslash$ scenario, and particularly for some choices of product
couplings considered above, these decays can take place at tree level
and hence their branching ratios could be substantially amplified over
their loop-driven SM predictions. As has been clarified in
\cite{Nardi}, the most general operators that contribute to the decay
of a neutral meson $M(= p\bar{q})$ into two charged leptons
$l\bar{l}$, could only be of the forms $c_P (\bar{q}\gamma_5
p)(\bar{l}\gamma_5 l)$, $c'_P (\bar{q}\gamma_5 p)(\bar{l} l)$ and $c_A
(\bar{q}\gamma_\mu \gamma_5 p)(\bar{l}\gamma_\mu \gamma_5 l)$.  In our
case, the branching ratio of $K_L$ decaying to $\mu^+\mu^-$ (tree level
$\tilde{u}_{Lj}$-exchanged decay) is given by
\begin{equation} 
B(K_L\rightarrow \mu^+\mu^-) =
\frac{(\lambda'_{2j1}\lambda'_{2j2})^2}{128\pi} \left(\frac{f_K
m_\mu}{m^2_{\tilde{u}_{Lj}}}\right)^2 \sqrt{1 - 4\frac{m_\mu^2}{m_K^2}}
\left(m_K \tau_{K_L}\right), 
\label{eq5}
\end{equation}
where $\tau_{K_L} (\simeq 5.17\times 10^{-8} s)$ is the
$K_L$-lifetime.  When the $\Rslash$ couplings are switched on, the
above branching ratio could be $\sim 217
(\lambda'_{2j2}.\lambda'_{2j1})^2$ for $m_{\tilde{u}_{Lj}} = 100$
GeV. Inserting the bound $\lambda'_{231}.\lambda'_{232} \ltap
7.7\times 10^{-4}$, obtained for $m_{\tilde{L}_e} = 100$ GeV, $B(K_L
\rightarrow \mu^+\mu^-)$ could touch $1.3\times 10^{-4}$ overshooting
the measured value ($(7.2 \pm 0.5) \times 10^{-9}$ \cite{pdg}) by four
or five orders of magnitude. So either we have to take
$m_{\tilde{t}_{L}} = 1.2$ TeV to compute the above branching ratio,
or, for $m_{\tilde{t}_{L}} = 100$ GeV, we obtain a more stringent
constraint $\lambda'_{231}.\lambda'_{232} \ltap 5.8\times
10^{-6}$. The process $B(K_L \rightarrow e^+e^-)$ is helicity
suppressed and from the experimental bound ($\ltap 4.1 \times
10^{-11}$ at 90\% CL \cite{pdg}) one obtains
$\lambda'_{1j1}.\lambda'_{1j2} \ltap 8.6\times 10^{-5}$ for
$m_{\tilde{u}_{Lj}} = 100$ GeV. From a similar consideration, $B(B_d
\rightarrow \tau^+\tau^-)$ could be estimated as $\sim
30(\lambda'_{3j1}.\lambda'_{3j3})^2$. With
$\lambda'_{311}.\lambda'_{313} \ltap 3.6 \times 10^{-3}$ (from $\Delta
m_{B_d}$), we infer that $B(B_d \rightarrow \tau^+\tau^-)$ could be as
high as $3.9\times 10^{-4}$ for a 100 GeV $\tilde{u}_L$. This can be
tested in future $B$-factories (the present experimental limit on this
branching ratio has been estimated \cite{Nardi} to be $1.5\times
10^{-2}$ from LEP data analysis).

Finally we turn our attention to the rare decay $K^+ \rightarrow \pi^+
\nu \bar{\nu}$, an evidence (only one event though) of which, citing a
branching ratio $B(K^+\rightarrow \pi^+\nu\bar{\nu}) =
4.2^{+9.7}_{-3.5}\times 10^{-10}$, has recently been reported by the
E787 Collaboration at BNL \cite{bnl}. The product couplings
$\lambda'_{ij1}.\lambda'_{ij2}$, constrained above by $K_L$--$K_S$
mixing and $K_L\rightarrow \mu^+\mu^-$ (or $e^+e^-$), drive this
interaction at tree level. The SM contribution \cite{BBL} is at least
one order of magnitude smaller than the experimental $1\sigma$ upper
limit ($1.4\times 10^{-9}$). Assuming the dominance of tree level
$\Rslash$ contribution, we obtain $B(K^+ \rightarrow \pi^+
\nu\bar{\nu}) = \left[\lambda'_{ij1}\lambda'_{ij2}/(4 G_F
m^2_{\tilde{d}_{Lj}} V_{us}^*)\right]^2 B(K^+ \rightarrow
\pi^0\nu\bar{e})$. Requiring then that it saturates the $1\sigma$
upper limit yields $\lambda'_{ij1}.\lambda'_{ij2} \ltap 1.6\times
10^{-5}$ for $m_{\tilde{d}_{Lj}} = 100$ GeV, where we have used
$B(K^+\rightarrow \pi^0\nu\bar{e}) = 0.0482$ \cite{pdg}. A look at
Table 1 reveals that for three combinations, corresponding to $i =$ 1,
2, or 3 and $j = 3$, the bounds are improved (indeed for a mass of 100
GeV of a different scalar). Turning the argument around, those three
product couplings individually are capable of reproducing the rare
event seen at BNL.

{\it To conclude}, we have derived new upper limits on several
combinations of $\lambda'$-couplings, product of two at a time,
by considering one-loop box graphs (with one $\tilde{L}_i$ and one
$W^\pm$/$G^\pm$/$H^\pm$ as internal lines) contributing to $\Delta
m_K$ or $\Delta m_B$. Most of our bounds are significantly tighter
than the previous ones. Meson decays to two charged leptons (in
particular, $K_L \rightarrow \mu^+\mu^-$ and $B_d \rightarrow
\tau^+\tau^-$) are enhanced in the presence of some of those product
couplings. Some could explain the rare $K^+ \rightarrow \pi^+
\nu\bar{\nu}$ event seen at BNL.

\vskip 5pt

Both authors thank the International Centre for Theoretical Physics,
Trieste, for its hospitality where the work was initiated. AR has
been supported in part by the Council of Scientific and Industrial
Research and the Department of Science and Technology, India.


\newpage 
\begin{table}[htbp]
\caption[] { New upper limits on product couplings for
$m_{\tilde{L}_i} = m_{H^\pm} = 100$ GeV. The combination marked with
``*'' (``**'') has a stronger upper limit as $8.6\times 10^{-5}$
($5.8\times 10^{-6}$) for $m_{\tilde{t}_L} = 100$ GeV (irrespective of
the charged Higgs boson mass), that follows from the consideration of
$K_L \rightarrow e^+e^- (\mu^+\mu^-)$. Similarly, the products marked
with `$\dagger$' have a stronger constraint $1.6\times 10^{-5}$ for
$m_{\tilde{b}_L} = 100$ GeV (again irrespective of the charged Higgs
boson mass), following from $K^+ \rightarrow \pi^+\nu \bar{\nu}$. }
\begin{center}
\vskip 5pt
\begin{tabular}{|c|c|c|c|c|c|c|}
\hline 
\hline

From & $\lambda'_{ijk}.\lambda'_{lmn}$ & Our limits & Previous
limits & $\lambda'_{ijk}.\lambda'_{lmn}$ & Our limits & Previous
limits \\

\hline

 & $(131).(132)[*,\dagger]$ & $7.7\times 10^{-4}$ & $1.2\times
10^{-2}$ & $(231).(232)[**,\dagger]$ & $7.7\times 10^{-4}$ &
$7.9\times 10^{-2}$ \\

 & $(331).(332)[\dagger]$ & $7.7\times 10^{-4}$ & $2.3\times 10^{-1}$
& $(131).(122)$ & $1.0\times 10^{-4}$ & $7.0\times 10^{-4}$ \\

 & $(231).(222)$ & $1.0\times 10^{-4}$ & $4.0\times 10^{-2}$ &
$(331).(322)$ & $1.0\times 10^{-4}$ & $9.6\times 10^{-2}$ \\

 & $(121).(122)$ & $1.4\times 10^{-6}$ & $7.0\times 10^{-4}$ &
$(221).(222)$ & $1.4\times 10^{-6}$ & $3.2\times 10^{-2}$ \\

 & $(321).(322)$ & $1.4\times 10^{-6}$ & $4.0\times 10^{-2}$ &
$(111).(112)$ & $1.4\times 10^{-6}$ & $7.0\times 10^{-6}$ \\

$\Delta m_K$ & $(211).(212)$ & $1.4\times 10^{-6}$ & $8.1\times
10^{-3}$ & $(311).(312)$ & $1.4\times 10^{-6}$ & $1.0\times 10^{-2}$
\\

 & $(122).(111)$ & $6.1\times 10^{-6}$ & $7.0\times 10^{-6}$ &
$(222).(211)$ & $6.1\times 10^{-6}$ & $1.6\times 10^{-2}$ \\

 & $(322).(311)$ & $6.1\times 10^{-6}$ & $2.0\times 10^{-2}$ &
$(132).(121)$ & $1.1\times 10^{-4}$ & $1.2\times 10^{-2}$ \\

 & $(232).(221)$ & $1.1\times 10^{-4}$ & $6.5\times 10^{-2}$ &
$(332).(321)$ & $1.1\times 10^{-4}$ & $9.6\times 10^{-2}$ \\

 & $(132).(111)$ & $4.7\times 10^{-4}$ & $1.2\times 10^{-4}$ &
$(232).(211)$ & $4.7\times 10^{-4}$ & $3.2\times 10^{-2}$ \\

 & $(332).(311)$ & $4.7\times 10^{-4}$ & $4.8\times 10^{-2}$ &
$(131).(112)$ & $2.4\times 10^{-5}$ & $7.0\times 10^{-4}$ \\

 & $(231).(212)$ & $2.4\times 10^{-5}$ & $2.0\times 10^{-2}$ &
$(331).(312)$ & $2.4\times 10^{-5}$ & $4.8\times 10^{-2}$ \\

\hline 

 & $(131).(133)$ & $1.3\times 10^{-3}$ & $2.4\times 10^{-5}$ &
$(231).(233)$ & $1.3\times 10^{-3}$ & $7.9\times 10^{-2}$ \\

 & $(331).(333)$ & $1.3\times 10^{-3}$ & $2.3\times 10^{-1}$ &
$(131).(123)$ & $1.8\times 10^{-4}$ & $7.0\times 10^{-3}$ \\

& $(231).(223)$ & $1.8\times 10^{-4}$ & $4.0\times 10^{-2}$ &
$(331).(323)$ & $1.8\times 10^{-4}$ & $9.6\times 10^{-2}$ \\

& $(111).(113)$ & $3.6\times 10^{-3}$ & $7.0\times 10^{-6}$ &
$(211).(213)$ & $3.6\times 10^{-3}$ & $8.1\times 10^{-3}$ \\

$\Delta m_{B_d}$ & $(311).(313)$ & $3.6\times 10^{-3}$ & $1.0\times
10^{-2}$ & $(121).(113)$ & $3.1\times 10^{-4}$ & $7.0\times 10^{-4}$
\\

& $(221).(213)$ & $3.1\times 10^{-4}$ & $1.6\times 10^{-2}$ &
$(321).(313)$ & $3.1\times 10^{-4}$ & $2.0\times 10^{-2}$ \\

& $(111).(123)$ & $1.6\times 10^{-2}$ & $7.0\times 10^{-5}$ &
$(211).(223)$ & $1.6\times 10^{-2}$ & $1.6\times 10^{-2}$ \\

& $(311).(323)$ & $1.6\times 10^{-2}$ & $2.0\times 10^{-2}$ &
$(121).(123)$ & $1.4\times 10^{-3}$ & $7.0\times 10^{-3}$ \\

& $(221).(223)$ & $1.4\times 10^{-3}$ & $3.2\times 10^{-2}$ &
$(321).(323)$ & $1.4\times 10^{-3}$ & $4.0\times 10^{-2}$ \\

\hline 

 & $(132).(133)$ & $2.5\times 10^{-3}$ & $2.4\times 10^{-4}$ &
$(232).(233)$ & $2.5\times 10^{-3}$ & $1.3\times 10^{-1}$ \\

 & $(332).(333)$ & $2.5\times 10^{-3}$ & $2.3\times 10^{-1}$ &
$(132).(113)$ & $1.5\times 10^{-3}$ & $6.8\times 10^{-3}$ \\

 & $(232).(213)$ & $1.5\times 10^{-3}$ & $3.2\times 10^{-2}$ &
$(332).(313)$ & $1.5\times 10^{-3}$ & $4.8\times 10^{-2}$ \\

 & $(112).(113)$ & $1.4\times 10^{-1}$ & $4.0\times 10^{-4}$ &
$(212).(213)$ & $1.4\times 10^{-1}$ & $8.1\times 10^{-3}$ \\

$\Delta m_{B_s}$ & $(312).(313)$ & $1.4\times 10^{-1}$ & $1.0\times
10^{-2}$ & $(122).(113)$ & $1.2\times 10^{-2}$ & $4.0\times 10^{-4}$
\\

 & $(222).(213)$ & $1.2\times 10^{-2}$ & $1.6\times 10^{-2}$ &
$(322).(313)$ & $1.2\times 10^{-2}$ & $2.0\times 10^{-2}$ \\

 & $(112).(123)$ & $3.2\times 10^{-2}$ & $4.0\times 10^{-3}$ &
$(212).(223)$ & $3.2\times 10^{-2}$ & $1.6\times 10^{-2}$ \\

 & $(312).(323)$ & $3.2\times 10^{-2}$ & $2.0\times 10^{-2}$ &
$(122).(123)$ & $2.7\times 10^{-3}$ & $4.0\times 10^{-3}$ \\

 & $(222).(223)$ & $2.7\times 10^{-3}$ & $3.2\times 10^{-2}$ &
$(322).(323)$ & $2.7\times 10^{-3}$ & $4.0\times 10^{-2}$ \\
\hline 
\hline
\end{tabular}
\end{center}
\end{table} 
\end{document}